\documentclass[letterpaper,hyper]{JHEP3}
\usepackage{epsfig}

\newcommand{\roughly}[1]{\mathrel{\raise.3ex\hbox{$#1$\kern-0.85em
\lower1ex\hbox{$\sim$}}}}

\newcommand{\lsim}{\roughly<}

\def\Expect#1{\langle #1 \rangle}

\def\cA{{\cal A}}

\def\cL{{\cal L}}
\def\cO{{\cal O}}

\def\exd{{\hbox{d}}}

\def\ba{\begin{eqnarray}}
\def\ea{\end{eqnarray}}
\def\be{\begin{equation}}
\def\ee{\end{equation}}

\def\ssB{{\scriptscriptstyle B}}
\def\ssC{{\scriptscriptstyle C}}
\def\ssE{{\scriptscriptstyle E}}

\def\ssL{{\scriptscriptstyle L}}

\def\ssR{{\scriptscriptstyle R}}

\def\IR{{\scriptscriptstyle I\kern-0.15em R}}
\def\SR{{\scriptscriptstyle SR}}
\def\EI{{\scriptscriptstyle EI}}
\def\BD{{\scriptscriptstyle BD}}
\def\HE{{\scriptscriptstyle HE}}

\def\ERo{E_R^{(2)}}

\def\bfk{{\bf k}}
\def\bfp{{\bf p}}

\def\O{\mathcal{O}}

\def\nn{\nonumber}

\def\({\left(}
\def\){\right)}

\def\pref#1{(\ref{#1})}

\def\Expect#1{\langle #1 \rangle}

\title{
Breakdown of Semiclassical Methods in de Sitter Space }

\author{C.P.~Burgess,${}^{1,2}$ R. Holman,${}^{3}$
L. Leblond${}^{2}$  and S. Shandera${}^{2}$\\
$^1$ Department of Physics and Astronomy,
  McMaster University, \\ \qquad \qquad Hamilton, Ontario, Canada;\\
$^2$ Perimeter Institute for Theoretical Physics,
  Waterloo, Ontario, Canada;\\
$^3$ Department of Physics, Carnegie Mellon University,\\
 \qquad \qquad Pittsburgh, Pennsylvania 15213;\\
 }

\date{}

\abstract { Massless interacting scalar fields in de Sitter space have long been known to experience large fluctuations over length scales larger than Hubble distances. A similar situation arises in condensed matter physics in the vicinity of a critical point, and in this better-understood situation these large fluctuations indicate the failure in this regime of mean-field methods. We argue that for non-Goldstone scalars in de Sitter space, these fluctuations can also be interpreted as signaling the complete breakdown of the semi-classical methods widely used throughout cosmology. By power-counting the infrared properties of Feynman graphs in de Sitter space we find that for a massive scalar interacting through a $\lambda \, \phi^4$ interaction, control over the loop approximation is lost for masses smaller than $m \simeq \sqrt \lambda \, H/2\pi$, where $H$ is the Hubble scale. We briefly discuss some potential implications for inflationary cosmology.}

\begin{document}

\section{Introduction}

Some aspects of quantum fields on de Sitter space remain
controversial, long after their first investigation more than 30
years ago \cite{dSorig}, but their potential relevance as an
explanation of the detailed properties of the fluctuations
observed by precision measurements \cite{CMBmeas, CMBth1} in the
Cosmic Microwave Background (CMB) radiation has stimulated a
recent re-examination of the issues \cite{Wbg, dSrecent1, dSrecent2,
KLR,AKH, Dolgov, MS, petri, us, SZ} (see also \cite{Review} and references therein)

The main interpretational difficulties for de Sitter space arise for massless fields, or for those that are very light compared with the Hubble scale: $m \ll H$. There are two related ingredients that complicate calculations with such fields: the presence of various types of infrared singularities, and the presence of large fluctuations over extra-Hubble distances.

For example, if a massless scalar field is prepared in an initial state for which fluctuations are small, then quantities like $\langle \phi^2(t) \rangle$ grow linearly with cosmic time, $t$. For massive scalars this growth eventually saturates at a $t$-independent value $\langle \phi^2 \rangle \propto H^4/m^2$, which is parametrically large if $m \ll H$ \cite{lint}. Since these fluctuations are uncorrelated, $\langle \phi(x) \phi(y) \rangle \simeq \langle \phi(x) \rangle \langle \phi(y) \rangle$, on scales longer than $H^{-1}$ and since gradients quickly redshift away, one finds a picture in which the field takes an approximately constant value within each Hubble volume, with different volumes evolving independently of one another in an essentially random and uncorrelated way \cite{superhubble}.

Although this has long been recognized as the appropriate physical picture, essentially all of what we know about fields in de Sitter backgrounds is based on calculations performed within the semi-classical approximation. This approximation assumes that classical field theory captures the dominant physics, with small calculable corrections arising from quantum fluctuations. The size of these corrections is believed to be kept small because of their
systematic dependence on small dimensionless quantities, like powers of coupling constants, $\lambda/(4\pi)^2$, and (because gravity is non-renormalizable) of small energy ratios, $E/M_p$ \cite{GREFT}. In particular, such an approximation underpins calculations of inflaton fluctuations during inflation, and their implications for the properties of the CMB at recombination.

In this paper we argue that for massless scalar fields in de Sitter backgrounds subject to non-derivative self-interactions --- like $V_{\rm int} \simeq \lambda \, \phi^4$ --- the presence of large extra-Hubble fluctuations undermines the entire semi-classical approximation. This is because the semiclassical approximation is at heart a mean-field approximation, within which a quantum field is represented as a dominant classical configuration plus a small quantum fluctuation, $\phi(x) = \varphi(x) + \delta\phi(x)$. But this kind of a description fundamentally breaks down over distances larger than $H^{-1}$ due to the large fluctuations occurring on these scales.

Notice that the breakdown of semiclassical methods we have in mind does not merely mean that the classical approximation is inadequate, with the situation being saved if we compute just a few more loops than usual. Instead, for massless scalars the danger is that the assumption that higher loops are suppressed by a small quantity that breaks down, meaning that {\em all} orders in the semiclassical expansion have similar sizes. This makes semiclassical calculations inherently unreliable because the truncation of the loop expansion omits contributions that are as large as those that are kept.

Quantum field theory is a cruel but fair master, so (as usual) the formalism contains within itself the news about the breakdown of semiclassical methods. The messenger is in this case the infrared divergences that commonly plague de Sitter calculations with massless scalars. These indicate a singular dependence on the mass in the more general case of a massive, but very light, scalar field. We argue that at least some of these divergences reflect the dominance of fluctuations over the contributions of the classical background, pointing to a fundamental breakdown of mean-field methods.

To make our claim precise, we consider a scalar field in de Sitter space, that self-interacts through a quartic scalar potential: $V = \frac12 \, m_0^2 \phi^2 + \frac{1}{4!} \, \lambda \, \phi^4$. Perturbing in $\lambda$ and using propagators for fields of mass $m_0$ shows that in the small-mass limit the usual loop-suppressing factor of $[\lambda/(2\pi)^2]$ for each loop in an $L$-loop graph is systematically modified by factors of $(H^2/m_0^2)$, indicating that higher loops are not suppressed once the scalar mass is sufficiently light.

We explicitly identify contributions to $L$-loop correlators that are proportional to
\be
 \left( \frac{\lambda H^4}{4 \pi^2 m_0^4} \right)^{L} \,,
\ee
which, taken at face value, would indicate that perturbation theory fails once $m_0^2 \sim \sqrt\lambda \, (H/2\pi)^2$. However we argue that the breakdown of perturbative methods at this higher mass arises because the physical mass scale that cuts off IR effects is really $m^2 = m_0^2 + \delta m^2$, with $\delta m^2 \simeq \lambda H^4/m_0^2$, and so it is only the expansion $H^2/m^2 = (H^2/m_0^2) [1 - \delta m^2/m_0^2 + \cdots]$ that breaks down when $m_0^2 \sim \sqrt\lambda \, (H/2\pi)^2$. This particular breakdown can be resummed: that is, it can be removed by reorganizing the perturbative expansion so that the unperturbed lagrangian involves the mass term $m^2 \phi^2$ rather than $m_0^2 \phi^2$. This particular reorganization does not also remove the potential breakdown of perturbation theory at $m_0^2 \simeq \lambda (H/2\pi)^2$.

Finite-temperature field theory provides a well-understood precedent for these conclusions. The small-momentum limit of the Bose-Einstein distribution function, $n_\ssB(k) \propto T/k$, implies that infrared divergences are stronger at finite temperature than they are at zero temperature. As a result, an $L$-loop Feynman graph for a self-interacting scalar field at finite temperature comes with the systematic factor $(\lambda \, T/4\pi^2 m_0)^{L}$, again indicating a breakdown of the loop expansion if $m_0 \lsim \lambda \,T/(2\pi)^2$. In the thermal case it is also known that the loss of perturbative control is only partial when $m_0 = 0$ since then the complete thermal mass is $m^2 =  \delta m^2 \propto \lambda T^2$ and the IR divergences can be weakened by reorganizing perturbation theory so that the unperturbed fields have this mass. In this case the same power-counting leads to the loop factor $\lambda T/m \propto \sqrt \lambda$, leading to controlled (but non-analytic) results for small $\lambda$. This same resummation fails, however, if $m_0^2 < 0$ is adjusted so that $m^2 = 0$, leading to a {\em bona fide} breakdown of semiclassical methods, such as is known to occur in condensed matter systems in the vicinity of a critical point. The breakdown of the loop expansion in this case underlies the well-known failure of mean-field methods to describe critical exponents \cite{RGCrit}. The physical origin of this breakdown is the dominance of large fluctuations near the critical point, similar to the fluctuations found in de Sitter space.

The analogy between thermal field theory and de Sitter space is robust, apart from one potentially important difference: in de Sitter space the difference $\delta m^2 = m^2 - m_0^2 \propto \lambda H^4/m_0^2$ itself depends singularly on $m_0^2$, unlike for thermal field theory where $\delta m^2 \propto \lambda T^2$. Because of this difference the map between the two cases is not simply to think of the Hubble scale as a temperature. In particular, although $m^2$ can be adjusted to vanish in the thermal case by appropriately choosing $m_0^2$, it is not clear that this can be done in the de Sitter situation. We discuss in the conclusions the open problem  of the extent to which these same regimes of resummation also apply in the de Sitter case (however, see \cite{Akhmedov:2008pu}).

Although gravity resembles a massless scalar in many ways, we
emphasize that we do {\em not} expect this same failure to arise
for pure gravity in a de Sitter background. The difference arises
because the gravitational self-interactions are largely derivative
couplings, and so are typically not as divergent in the infrared.
Since Goldstone scalars similarly couple derivatively, they also need
not share the same mean-field breakdown. Massless spin-1
fields can couple without derivatives, and infrared effects are
also known to ruin mean-field methods for a hot plasma of charged
particles interacting through gauge interactions \cite{HTL}. We
leave it open whether a similar breakdown occurs on de Sitter
space, but any such a failure would require the existence of very
light charged degrees of freedom to survive the exponential de
Sitter red-shifting. There is some work for SQED in de Sitter space that shows that the photon can indeed get a mass there (see \cite{PW} for a review).

The remainder of the paper is organized as follows. The next
section, \S2, briefly reviews the power-counting of infrared
divergences for self-interacting scalar fields at finite
temperature, to show how these track the breakdown of mean-field
methods. \S3 then provides a similar power-counting for
self-interacting scalars in de Sitter space. Dimensional analysis
is first used to argue that the worst divergences are logarithmic for
any correlation functions. For a massive propagator, this translates into inverse powers of mass at each loop order. A class of graphs is then displayed that verifies this dependence through explicit calculation. Finally, \S4\ draws some preliminary implications for inflationary
calculations, and summarizes our conclusions.

\section{IR divergences at finite temperature}

We first review the breakdown of mean-field methods near a
critical point. To this end we estimate the size of the
contribution to physical amplitudes of the infrared divergences
that arise due to long-wavelength fluctuations. The purpose is to
sketch why these long-wavelength fluctuations eventually dominate
the suppressions due to small couplings at any order in a
perturbative expansion.

Consider then a massive scalar field with a quartic self
interaction in flat spacetime
\be \label{lagr0}
 \cL = - \frac12 \, \partial_\mu \phi \, \partial^\mu \phi
 - \frac{m_0^2}{2} \, \phi^2 - \frac{\lambda}{4!} \, \phi^4 \,,
\ee
heated to a nonzero temperature $T$. In Euclidean signature the
scalar propagator is
\be
 G_n(\bfp) = - \frac{i}{(2\pi)^3} \; \frac{1}{p_n^2 + \bfp^2 + m_0^2}
 \,,
\ee
where $p_n = 2\pi n/\beta = 2\pi nT$, with $n = 0, \pm 1, \pm 2,
\cdots$, ensures periodicity in imaginary time $\tau \to \tau +
\beta$.

Consider now a Feynman graph that involves $E$ external lines, $I$
internal lines and $V$ vertices, and (to start) suppose there are no insertions of the mass counter-term vertex $\delta m^2$. Then the parameters $E$, $I$ and $V$ are related by the identity (conservation of ends) $E + 2I = 4V$ and the definition of the number of loops, $L = I - V + 1$. Using these to eliminate $I$ and $V$ then gives
\be
 2V = E + 2(L - 1) \quad \hbox{and} \quad
 2I = E + 4(L - 1) \,.
\ee

If $k \simeq m_0$ denotes a typical external momentum scale, the
resulting amplitude is schematically given by
\be
 \cA^0(\bfk, T) \simeq  \left[ \lambda  (2\pi)^3
 \delta^3(\bfp + \bfk) \, \frac{1}{T} \, \delta_{nn'} \right]^V
 \left[T \sum_n  \int  \frac{\exd^3 \bfp }{(2\pi)^3}
 \; \frac{1}{p_n^2 + \bfp^2 + \bfk^2 + m_0^2} \right]^I \,.
\ee
All of the delta functions perform one of the integrals, except
for one which expresses the overall momentum conservation delta
function.

The most IR singular part of the result corresponds to the $n = 0$
term of the sums, so if the IR convergence occurs at scales $p
\simeq m_0$ (more about this below) then for $k \simeq m_0 \ll T$ the most singular dependence on $m_0$ has the form
\ba \label{PChiT0}
 \cA^0_\IR(\bfk, T) &\simeq&  \delta^3(\bfk) \; \frac{\lambda^{V}}{T}
 \left[ T  \int  \frac{\exd^3 \bfp }{(2\pi)^3} \right]^{I-V+1}
 \left[ \frac{1}{\bfp^2 + \bfk^2 + m_0^2} \right]^I \nn\\
 &\simeq&  \delta^3(\bfk) \; \frac{\lambda^{L - 1 + E/2}}{T}
 \left[T \int  \frac{\exd^3 \bfp }{(2\pi)^3} \right]^{L}
 \left[ \frac{1}{\bfp^2 + \bfk^2 + m_0^2} \right]^{2L - 2 + E/2} \nn\\
 &\simeq&  \delta^3(\bfk) \; \lambda^{L - 1 + E/2} T^{L - 1}
 \left[ \frac{2\pi}{(2\pi)^3} \right]^{L}
  m_0^{3L - 4L + 4 - E}  \nn\\
 &\simeq&  \delta^3(\bfk) \, \left( \frac{m_0^4}{\lambda T} \right)
  \left( \frac{\lambda}{m_0^2} \right)^{E/2}
 \left[ \frac{\lambda \, T}{(2\pi)^2 m_0} \right]^{L} \,.
\ea
This is the main result, expressing the most singular small-$m_0$
limit of an $L$-loop contribution.

Perhaps the biggest surprise in the estimate \pref{PChiT0} is the systematic appearance (when $m_0 \ll T$) of the factor $T/m_0$ at each loop order. Once $m_0^2 \lsim \lambda^2 T^2/16 \pi^4$ this eventually undermines the semiclassical loop expansion itself, whose validity ultimately rests on the suppression of loops by powers of $\lambda \ll 1$.

What happens in the limit $m_0^2 \to 0$ can be understood by reorganizing perturbation theory to recognize that finite-temperature effects also contribute to the scalar mass, and so can themselves suppress the total failure of an expansion in powers of $\lambda$. This reorganization can be made explicit by rewriting the lagrangian, eq.~\pref{lagr0}, as
\be
 \cL = - \frac12 \, \partial_\mu \phi \, \partial^\mu \phi
 - \frac{m^2}{2} \, \phi^2 + \frac{\delta m^2}{2} \, \phi^2 - \frac{\lambda}{4!} \, \phi^4 \,,
\ee
where $m^2 = m_0^2 + \delta m^2$ is the physical scale responsible for saturating IR fluctuations, and $\delta m^2$ is a perturbatively small, but calculable, thermal mass shift, whose leading form is $\delta m^2 \simeq c \lambda T^2/4\pi^2$ for $c$ a positive constant of  order unity.

Now consider repeating the above power-counting argument, but also inserting $n$ factors of the mass counter-term, $\delta m^2$, into the result. Since each such insertion also adds a new propagator, it contributes an amount $\sim \delta m^2/\bfp^2$, so the same dimensional argument as given above for the leading IR behaviour results in the revised estimate
\be \label{PChiT}
 \cA^n_\IR(\bfk, T) \simeq  \delta^3(\bfk) \,
 \left( \frac{m^4}{\lambda T} \right)
  \left( \frac{\lambda}{m^2} \right)^{E/2}
 \left[ \frac{\lambda \, T}{(2\pi)^2 m} \right]^{L}
 \left( \frac{\delta m^2}{m^2} \right)^n \,.
\ee
In the limit where the zero-temperature mass vanishes, $m_0^2 = 0$, then $m^2 = \delta m^2 \simeq c \, \lambda\, T^2/4\pi^2$, and so all of the factors $\delta m^2/m^2$ are order one. But the point of this reorganization is that these new order-unity interactions proportional to $\delta m^2$ systematically cancel order-unity parts of the self-energy corrections in Feynman graphs, leaving a result that is suppressed by a net power of $\lambda$.

For instance, the one-particle reducible contribution to the 2-point function at zero external momentum at two loops has an IR singularity proportional to $\lambda^2 T^4/m^2$. But this precisely cancels the contributions of the one-particle reducible graphs involving $\delta m^2$ insertions, to leave a residual contribution that is of order $\lambda^2 T^3/m \propto \lambda^{3/2} T^2$, in agreement with standard calculations. In general, using $\delta m^2 = m^2 \simeq \lambda \, (T/2\pi)^2$ in eq.~\pref{PChiT} shows that the small parameter that suppresses each loop in this case is
\be
 \left( \frac{\lambda \, T}{(2\pi)^2 m} \right)^{L} \simeq
 \left( \frac{\sqrt\lambda \;}{2\pi} \right)^{L} \,.
\ee
The contribution of higher-loop graphs remain small for small $\lambda$, although each additional loop now costs a factor of $\sqrt\lambda$ rather than $\lambda$ itself. This initially surprising non-analytic dependence on the coupling $\lambda$ can arise within perturbation theory because of the
reorganization of perturbation theory that is implicit in using the loop-generated mass to cut off the infrared divergences. For scalars it is known that the shift in mass completely resolves the apparent breakdown of perturbation theory \cite{HTL}, though the same is not true for massless gauge bosons in hot plasmas.

A situation where the above resummation does {\em not} salvage the semiclassical loop expansion is when the zero-temperature mass is negative and adjusted in precisely the way required to ensure that the full
finite-temperature mass vanishes for a particular temperature: $m_0^2 = - \delta m^2 \simeq - c \lambda T^2/4\pi^2$, so that $m^2 \simeq 0$. This is what happens at a critical point, $T = T_c$, such as appears at the termination of a phase-coexistence curve. In such a case the vanishing of $m$ in the estimate \pref{PChiT} indicates that control over the loop expansion breaks down completely, in agreement with the well-known failure of mean-field methods to compute critical exponents near critical points (where $m \to 0$) \cite{RGCrit}. The root of the problem with the loop expansion in this case lies in its mean-field nature, since the large fluctuations allowed by the massless fluctuations at a critical point invalidate the expansion about a large background field.

\section{IR divergences in dS space}

We next repeat the above power-counting argument for calculations
of scalar-field correlations in a fixed background de Sitter
spacetime. To this end we work within the `in-in' formalism
\cite{inin}, and imagine computing a correlation function for
$\phi(x)$ for a minimally coupled scalar field that self-interacts
through the potential
\be
 V(\phi) = V_0 + \frac{m_0^2}{2} \, \phi^2 + \frac{\lambda}{24} \,
 \phi^4   \,.
\ee
We take the background cosmological constant, $V_0$, to be
sufficiently large as to dominate the quantum fluctuations of
$\phi$, allowing the background curvature to be regarded as a
fixed geometry $\exd s^2 = a^2(\tau) \left( - \exd \tau^2 +
\delta_{ij} \exd x^i \exd x^j \right)$, where $\tau$ is conformal
time and the de Sitter scale factor is $a = e^{H t} = -1/(H
\tau)$.

\subsection{In-in correlations}

Within the in-in formalism the path integration is over a
duplicate set of field configurations, $\phi_+$ and $\phi_-$, that
correspond to the two time-paths to temporal infinity that arise,
but it is useful to instead use the two combinations $\phi_\ssC =
\frac12(\phi_+ + \phi_-)$ and $\phi_\Delta = \phi_+ - \phi_-$.
Since the path integral is weighted by the integrand $\exp\left[ i
S(\phi_+) - i S(\phi_-) \right]$, in terms of the fields
$\phi_\ssC$ and $\phi_\Delta$ the scalar self-interaction
appearing within the path integral therefore has the form
\ba
 V(\phi_+) - V(\phi_-) &=& \frac{m_0^2}{2} (\phi_+^2 - \phi_-^2)
 + \frac{\lambda}{24} (\phi_+^4 - \phi_-^4) \nn\\
 &=& m_0^2 \phi_\ssC \phi_\Delta
 + \frac{\lambda}{24} \Bigl( 4\phi_\ssC^3 \phi_\Delta
 + \phi_\ssC \phi_\Delta^3 \Bigr) \,,
\ea
so these are the interactions whose vertices we follow in any
particular Feynman graph.

The internal lines of the graph represent the correlations of
these fields: $\langle \phi_\ssC \phi_\ssC \rangle$, $\langle
\phi_\Delta \phi_\ssC \rangle$ and $\langle \phi_\ssC \phi_\Delta
\rangle$ (part of the utility of the combinations $\phi_\ssC$ and
$\phi_\Delta$ is the vanishing of the autocorrelation $\langle
\phi_\Delta \phi_\Delta \rangle = 0$). The correlator $\langle
\phi_\ssC (\tau_1) \phi_\ssC (\tau_2) \rangle$ for a massless
scalar propagator becomes
\ba\label{BDmassless}
 G_\ssC^0(k,\tau_1,\tau_2)
  &=& \frac{H^2}{2k^3} \, \Bigl\{ (1 + k^2 \tau_1 \tau_2)
 \cos[ k(\tau_1 - \tau_2)] + k(\tau_1 - \tau_2) \sin [
 k (\tau_1 - \tau_2)] \Bigr\} \,, \nn\\
 &\simeq& \frac{H^2}{2 k^3} \Bigl\{ 1 + \cO[(k\tau_i)^2] \Bigr\}
 \,,
\ea
where the last line specializes to the long-wavelength,
super-Hubble limit,
\be
 -k \tau = \frac{k}{aH} \ll 1 \,.
\ee
The retarded correlator, $\langle \phi_\ssC(\tau_1) \phi_\Delta
(\tau_2) \rangle$, is similarly
\ba
 G_\ssR^0(k, \tau_1,\tau_2)
 &=& \theta(\tau_1 - \tau_2) \,
 \frac{H^2}{k^3} \, \Bigl\{ (1 + k^2 \tau_1 \tau_2)
 \sin[ k(\tau_1 - \tau_2)] - k(\tau_1 - \tau_2) \cos [
 k (\tau_1 - \tau_2)] \Bigr\} \nn\\
 &\simeq& \theta(\tau_1 - \tau_2) \,
 \frac{H^2}{3} (\tau_1^3 - \tau_2^3)
 \Bigl\{ 1 + \cO\left[(k\tau_i)^2 \right] \Bigr\}  \,.
\ea

\subsection{Power-counting}

For the purposes of power-counting imagine computing a correlation
function involving $N_\ssC$ powers of $\phi_\ssC$ and $N_\Delta$
powers of $\phi_\Delta$: $\Expect{\phi_{\Delta}^{N_\Delta}
\phi_\ssC^{N_\ssC}}$. A generic Feynman graph contributing to such
a quantity involves $I_R$ internal lines involving retarded (or
advanced) propagators, $\Expect{\phi_\ssC \phi_\Delta}$ and
$I_\ssC$ lines representing $\Expect{\phi_\ssC \phi_\ssC}$
propagators; linking $V_\ssC$ vertices describing the $\phi_\ssC^3
\phi_\Delta$ interaction, and $V_\Delta$ vertices built from
$\phi_\ssC \phi_\Delta^3$ --- see Figure (\ref{fig:vertices}). If we write $m_0^2 = m^2 - \delta m^2$ then graphs also involving insertions of $\delta m^2$ have additional factors, but for the moment we ignore these.

\FIGURE[ht]{ \epsfig{file=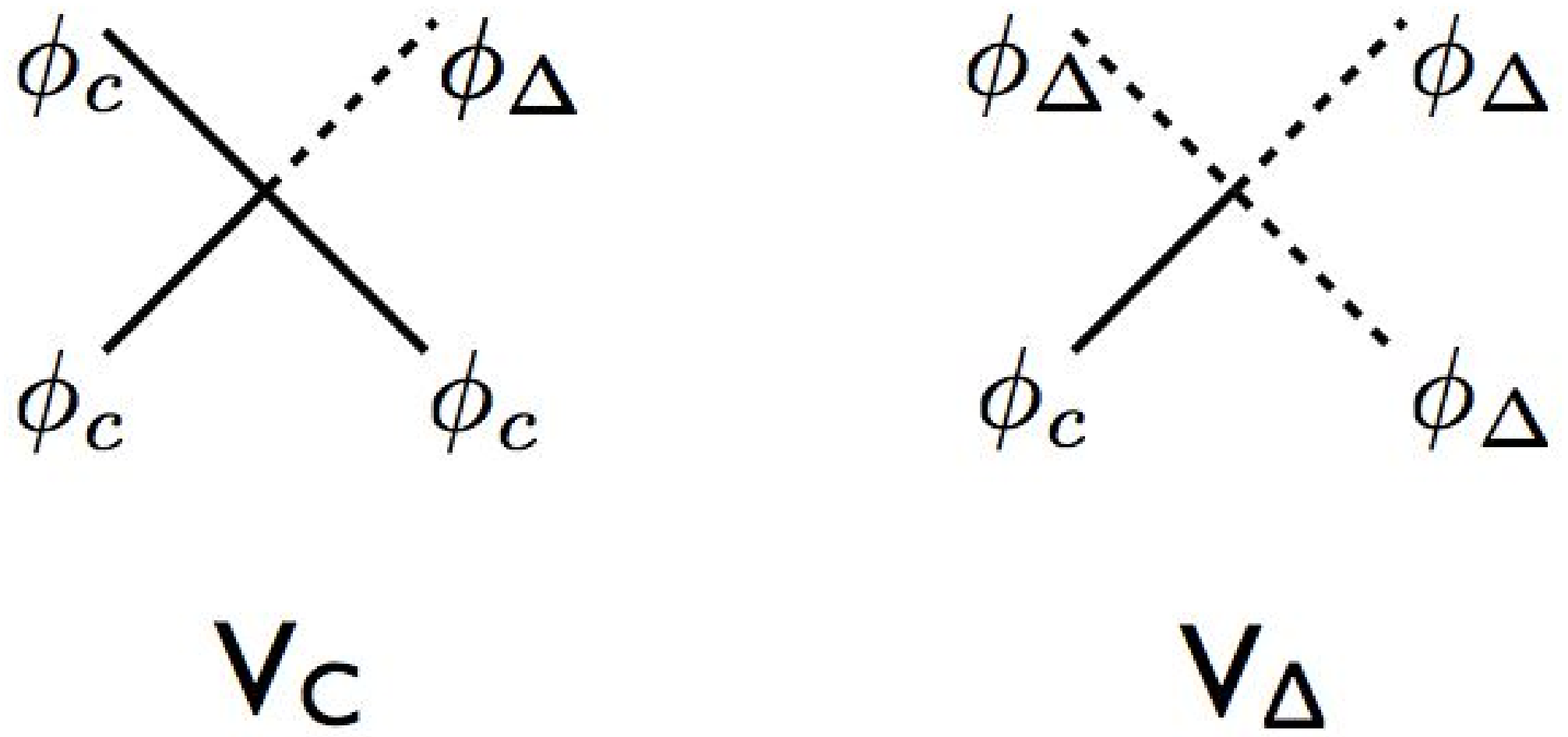,angle=0,width=0.4\hsize}
\qquad \epsfig{file=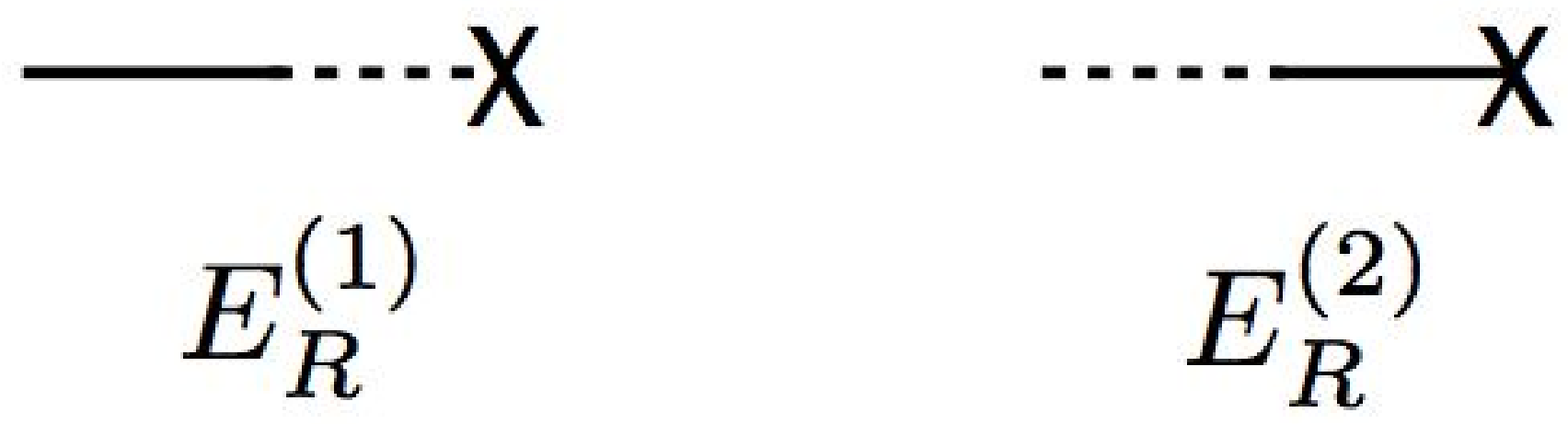,angle=0,width=0.4\hsize}
\caption{Labelling for vertices and external lines. For external
lines the {\it X} indicates which end attaches to the vertex.}
\label{fig:vertices} }

There are also $E_\ssC$ external lines corresponding to the
$G^0_\ssC$ propagators, each of which must attach to one of the
$N_\ssC$ $\phi_\ssC$ fields in the correlation. Due to the
off-diagonal nature of the external retarded propagators these can
connect to either an external $\phi_\ssC$ or $\phi_\Delta$ field.
If there are $E_\ssR$ such external propagators we denote by
$E_\ssR^{(1)}$ (respectively $E_\ssR^{(2)}$) the number of such
propagators that connect to an external $\phi_\ssC$ (respectively
$\phi_\Delta$) field, as shown in Figure (\ref{fig:vertices}).
Clearly $E_\ssR^{(1)}+E_\ssC= N_\ssC$ and $E_\ssR^{(2)} =
N_\Delta$, and so $E_\ssR^{(1)} + E_\ssR^{(2)} + E_\ssC = E_\ssR +
E_\ssC = N_\ssC + N_\Delta$.

The numbers of propagators and vertices are related by the
definition of the number of loops
\be
 L-(I_\ssC+I_\ssR)+(V_\ssC+V_{\Delta})=1
\ee
and the two conditions expressing the conservation of `$C$' and
`$\Delta$' type ends,
\ba \label{rules1}
 E_\ssC + 2I_\ssC + I_\ssR + E_\ssR^{(2)}
 &=& 3V_\ssC + V_{\Delta} \nn\\
 I_\ssR + E_\ssR^{(1)} &=& V_\ssC + 3V_{\Delta} \,.
\ea
These last two use the fact that `$R$' type propagators (involving
retarded or advanced Greens functions) are mixed correlations that
have one $\phi_\ssC$ end and one $\phi_\Delta$ end, while `$C$'
type propagators involve $\phi_\ssC$ fields at both ends. It is
useful to use these three identities to eliminate $V_\ssC$,
$I_\ssR$ and $I_\ssC$, to get
\ba \label{IV1}
 I_\ssR &=& \frac12 \left( E_\ssR^{(2)}
 + E_\ssC - E_\ssR^{(1)} \right) + L - 1 + 2V_{\Delta} \nn\\
 I_\ssC &=& E_\ssR^{(1)} + L - 1 - 2V_{\Delta} \\
 V_\ssC &=& \frac12 \left( E_\ssR^{(2)} + E_\ssC + E_\ssR^{(1)} \right) + L - 1 - V_{\Delta} \,. \nn
\ea

Now in a generic Feynman graph, each vertex contributes a factor
that has the schematic form
\ba
 \hbox{Vertex} &\simeq& \lambda \int \frac{\exd \tau}{H^4\tau^4}
 \, (2\pi)^3 \delta^3(p) \nn\\
 \ea
where $\tau$ is the time that labels the vertex and the delta
function expresses conservation of the (co-moving) momenta, $p$,
that meet at the vertex. Similarly, for massless scalars each internal `$C$' type propagator contributes in the IR limit
\be
 \hbox{$C$-type propagator} \simeq \int \frac{\exd^3
 p}{(2\pi)^3} \; \frac{H^2}{p^3} \,,
\ee
while each internal `$R$' type propagator gives
\ba
 \hbox{$R$-type propagator} &\simeq& \int \frac{\exd^3
 p}{(2\pi)^3} \; H^2 \theta(\tau_i - \tau_j) \left(
 \tau_i^3 - \tau_j^3 \right) \,.
\ea

Combining results, evaluating a Feynman graph gives a result
proportional to
\be
 A(k_1,\tau_1; \dots ; k_\ssE, \tau_\ssE)
 \simeq  (2\pi)^3 \delta^3(k) \cA(k_1, \tau_1;
 \dots ; k_\ssE, \tau_\ssE) \,,
\ee
where the delta function schematically represents the overall
conservation of spatial momentum. The expression $\mathcal{A}$
is given schematically by
\ba\label{powergeneral}
 \mathcal{A}(k,\tau_a) &\propto& H^{2(I_\ssC+I_\ssR)}
 \left( \frac{H^2}{2k^3} \right)^{E_\ssC}
 \left[ \lambda \int \frac{\exd\tau_i}{H^4\tau_i^4}
 \right]^{V_\ssC + V_{\Delta}} \left[
 H^2 \theta (\tau_a - \tau_i) (\tau_a^3 -
 \tau^{3}_i) \right]^{E_\ssR^{(1)} + E_\ssR^{(2)}} \nonumber\\
 &\times& \left[ \int \frac{\exd^3p}{(2\pi)^3}
 \frac{1}{2p^3} \right]^{I_\ssC} \left[
 \int \frac{\exd^3p}{(2\pi)^3} \frac{ \theta( \tau_j
 - \tau_l)}{3} (\tau_j^3 - \tau_l^{3})
 \right]^{I_\ssR} \left[ (2\pi)^3 \delta^3(p)
 \right]^{V_\ssC + V_{\Delta} - 1}
\ea
where the $k$'s generically stand for external momenta, the $p$'s for loop momenta, and we use the early part of alphabet ($a,b,c$) for external time and the middle part $i,j,l$ for internal times associated with vertices. In the Appendix we examine several Feynman graphs explicitly, to examine the behaviour of the integrations over the vertex times, $\tau_i$, and loop momenta, $p$.

\subsection{IR behaviour}

We next estimate the most IR-sensitive part of such a de Sitter-space graph. For technical simplicity, even though we imagine the IR regulator to come from nonzero $m^2$, we perform our estimate by working with $m_0^2 = 0$ and cutting off the divergence at a physical scale $\Lambda_\IR$, relating this scale to $m_0$ and $H$ in a second step.

The dependence of the final result, $\cA$, on both momenta, $k$,
and external times, $\tau_a$, complicates the use of dimensional
analysis in identifying the dominant infrared divergences that can
arise in general amplitudes. To make progress we assume the fields
being correlated are evaluated in position space, in which case
the position-space amplitudes are obtained from the above
expressions by multiplying by an appropriate factor of
$(2\pi)^{-3} \int \exd^3 k \; e^{i k \cdot x}$. Power-counting
then simplifies in the special case that all external fields are
evaluated at the same position, since then de Sitter invariance of
the vacuum implies $\Expect{\phi^{N_\ssC}_\ssC(x)
\phi^{N_\Delta}_\Delta(x)}$ is independent of $x^\mu$. In this
case the result cannot depend on $k$ or $\tau_a$, and so the
dominant divergence must be a function only of $H$ and the IR
cutoff, $\Lambda_\IR$. Notice that this argument requires
$\Lambda_\IR$ to be a time-independent cutoff on physical momenta,
since a time-independent cutoff on co-moving momentum would
introduce a spurious time-dependence into the correlation
function. (In this respect we follow \cite{us}, and at least for
the UV \cite{SZ}, but differ from the practice used by other
recent workers.)

Given the result, eq.~\pref{powergeneral}, of the last section
(and after UV divergences are renormalized by including the
appropriate counterterm graphs) the dominant IR divergence of a
position-space amplitude of the above type can be determined by
dimensional analysis
\ba \label{powergeneralposn}
 \Expect{\phi_\ssC^{N_\ssC}(x) \, \phi_\Delta^{N_\Delta}(x)}
 &\sim& \left[ \int \frac{\exd^3 k}{(2\pi)^3}
 \right]^{N_\ssC + N_\Delta} \Bigl[ (2 \pi)^3
 \delta^3(k) \Bigr] \; \cA(k,\tau_a) \nonumber\\
 &\propto& H^{2(I_\ssC+I_\ssR +E_\ssC + E_\ssR)
 - 4( V_\ssC + V_\Delta)} \lambda^{(V_\ssC + V_\Delta)} \nn\\
 && \quad \times
 \left[ \int \frac{\exd^3 k}{(2\pi)^3}
 \right]^{N_\ssC + N_\Delta} \left(
 \frac{1}{k^3} \right)^{E_\ssC}
 \left[ (2\pi)^3 \delta^3(k) \right] \nn\\
 && \quad\quad \times \left[ \int
 \frac{\exd^3 p}{(2\pi)^3} \right]^{I_\ssC + I_\ssR}
 \left[\frac{1}{p^3} \right]^{I_\ssC} \left[
 (2\pi)^3 \delta^3(p) \right]^{V_\ssC
 + V_{\Delta} - 1} \nonumber\\
 && \quad\quad\quad \times  \left[ \int
 \frac{\exd\tau_i}{\tau_i^4} \right]^{V_\ssC + V_{\Delta}}
 \left[ \theta (\tau_a - \tau_i) (\tau_a^3 - \tau^{3}_i)
 \right]^{E_\ssR} \left[ \theta( \tau_j - \tau_l)
 (\tau_j^3 - \tau_l^{3}) \right]^{I_\ssR}  \nn\\
  &\propto& \left[ \frac{\sqrt\lambda \; H}{(2 \pi)^2}
  \right]^{N_\ssC + N_\Delta} \left( \frac{\lambda}{4\pi^2}
  \right)^{L-1} \Lambda_\IR^{3P} \,,
\ea
with
\be
 P = (N_\ssC + N_\Delta) - E_\ssC - 1 +
 (I_\ssC + I_\ssR) - I_\ssC - (V_\ssC + V_\Delta -1)
 + (V_\ssC + V_\Delta) -E_\ssR - I_\ssR  = 0 \,.
\ee
Since $P = 0$, we find that in general every diagram contributing
to this type of correlation function is at worst log divergent in
the infrared. Of course, the above dimensional argument cannot in itself distinguish a divergence like $\ln \Lambda_\IR$ from $(\ln \Lambda_\IR)^L$, and so a more precise determination (given below) of the nature of the divergence requires a more detailed estimate.

\subsection{Massive Propagators}

To better parse how this divergence is regulated as a function of
the scalar mass, we step back and use the small-$k$ behaviour that
is relevant to the massive scalar propagator on de Sitter space.
In the limit $-k\tau = k/(aH) \ll 1$ this becomes
\ba\label{massive}
 G_\ssC(k,\tau_1,\tau_2) & \simeq &
 \frac{H^2}{2k^3} (k^2\tau_1\tau_2)^{\epsilon_0}\\
 G_\ssR(k,\tau_1,\tau_2)  & \simeq & \theta(\tau_1-\tau_2)
 \frac{H^2}{3}(\tau_1^{3-\epsilon_0} \tau_2^{\epsilon_0} -
 \tau_1^{\epsilon_0} \tau_2^{3-\epsilon_0}) \,,
\ea
where $\epsilon_0 = {m_0^2}/{3H^2}$. Even though these expressions
differ only weakly for small $k$ from the massless case, they
suffice to cure the IR divergences encountered previously because
these divergences are only logarithmic.

The main change that this introduces relative to the above
estimates is the conversion of the internal-line factor
\be
 \left[ \frac{H^2}{p^3} \right]^{I_\ssC}
 \to \left[ \frac 1H \left( \frac{H}{p} \right)^{3 - 2\epsilon_0}
  \right]^{I_\ssC} \,,
\ee
which, using as before a cutoff, $\Lambda_\IR$, on physical
momenta gives
\ba
 \int_{\Lambda_\IR}^P \frac{\exd p}{p} \; \left( \frac{H}{p}
 \right)^{-2\epsilon_0}
 &\simeq&  \ln \left( \frac{P}{\Lambda_\IR} \right)
 \qquad\qquad\qquad
 \hbox{if $\epsilon_0 \to 0$} \nn\\
 &\simeq& \frac{3H^2}{2 m_0^2} \left( \frac{P}{H}
 \right)^{2 m_0^2/3H^2} \qquad \hbox{if $\Lambda_\IR \to 0$} \,.
\ea
This shows that the small-mass limit converts powers of $\ln
\Lambda_\IR$ into powers of $H^2/m_0^2$. This conversion of a
logarithmic divergence into an inverse power of $m_0$ (as opposed to
powers of $\ln m_0$) arises because for $m_0^2 \ll H^2$ the difference
between the massive and massless expressions for $G_\ssC(k,\tau)$
becomes important only once $(-k \tau )^{2\epsilon_0}$ deviates from
unity, which occurs for $k < k_*$ with
\be
 -k_* \tau = \frac{k_*}{aH} \simeq e^{-1/2\epsilon_0} = e^{-3H^2/2m_0^2}
 \,,
\ee
corresponding to a physical IR cutoff of order $\Lambda_{\IR} \simeq k_*/a
\simeq H \, e^{-3H^2/2m_0^2}$.

\subsection{Powers of logs}

Because power-counting returned the result that graphs diverge like $\Lambda_\IR^P$ in the infrared with $P=0$, we know these divergences are logarithmic. What these arguments do not yet show is that this logarithmic divergence worsens order-by-order in the loop expansion. To establish this we next identify an explicit class of graphs for which additional powers of $H/m$ arise at each order in perturbation theory.

\FIGURE[ht]{ \epsfig{file=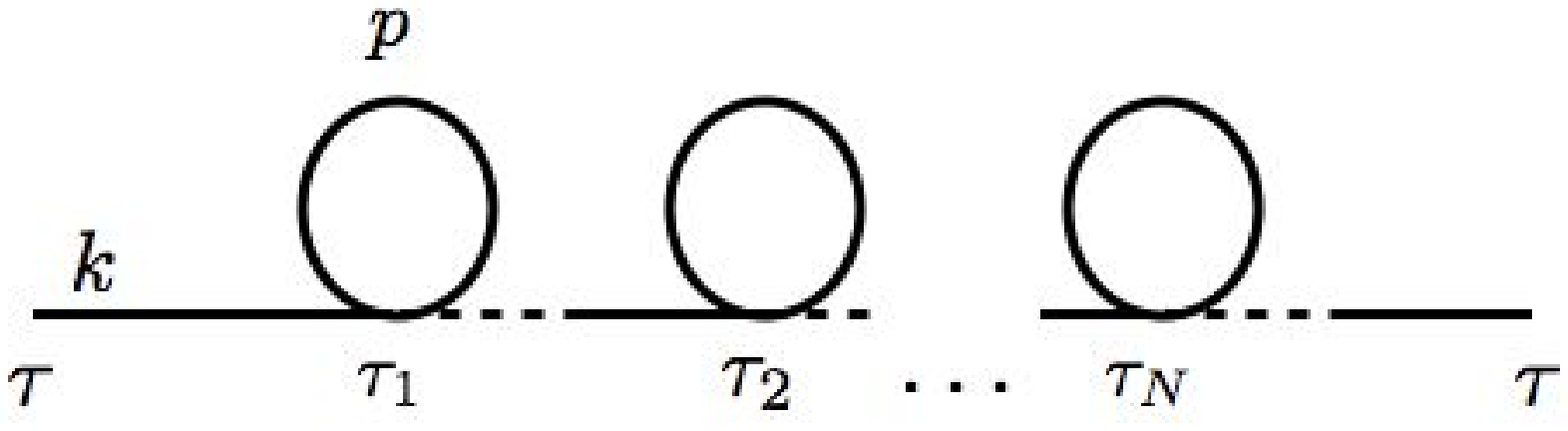,angle=0,width=0.6\hsize}
\caption{A class of graphs contributing terms of order
$(\lambda H^2/m_0^2)^L$ to $G_\ssC(k,\tau)$.} \label{fig:chain} }

The explicit class of graphs we choose for this purpose are those containing the successive chain graphs of Fig.~\ref{fig:chain}. These diagrams are loop corrections to $G_\ssC(k,\tau) = \Expect{\phi_\ssC^2}$ and we denote by $G_\ssC^{(\ssL)}$ the $L$-loop chain diagram contribution to $\Expect{\phi_\ssC^2}$. The 1-loop contribution, $G_\ssC^{(1)}$, is evaluated in great detail in \cite{petri,us} so we just quote the end result in momentum space:
\be
 G_\ssC^{(1)}(k,\tau) = -\lambda \int_{-\infty}^0
 \exd\tau'a^4(\tau') G_\ssC^{(0)}(k,\tau,\tau')
 G_\ssR^{(0)}(k,\tau',\tau) \Lambda(\tau')
\ee
where
\be
 \Lambda(\tau) = \int\frac{\exd^3p}{(2\pi)^3} \;
 G_\ssC^{(0)}(k,\tau,\tau)
\ee
is the loop factor. With the appropriate (ultraviolet) mass counterterm and
using the massive propagator given by Eq.~(\ref{massive}), one
finds that the loop factor is time independent and equal to
\cite{us}
\be
 \Lambda(\tau) = \frac{1}{2\epsilon_0} \left(
 \frac{H}{2\pi} \right)^2 \left(\frac{\mu}{H}
 \right)^{2\epsilon_0} \,,
\ee
where $\mu$ is some non-IR physical scale. All together, the
1-loop correction to $G_\ssC$ is
\be
 G_\ssC^{(1)}(k,\tau,\tau)) = G_\ssC^{(0)}(k,\tau,\tau)
 \frac{\lambda}{6(2\pi)^2\epsilon_0}
 \left(\frac{\mu}{H}\right)^{2\epsilon_0} \ln(-k\tau) \,.
\ee
At $L$ loops the chain diagrams simply factorize, so that the
$L$-loop chain diagram is just
\be
 G_\ssC^{(\ssL)}(k,\tau,\tau) \propto G_\ssC^{(0)}
 (k,\tau,\tau) \left[ \frac{\lambda}{(2\pi)^2\epsilon_0}
 \left( \frac{\mu}{H}\right)^{2\epsilon_0}
 \ln(-k\tau)\right]^L \,.
\ee
Combining the factors depending on $L$ then shows that each
successive loop comes systematically pre-multiplied by a factor of
\be
 \left( \frac{\lambda H^2}{4 \pi^2 m_0^2} \right)^{L} \,,
\ee
indicating the breakdown of the loop expansion once $m_0^2$ is sufficiently small.

To make contact with our power-counting estimate above we Fourier transform $G_\ssC^{(\ssL)}(k,\tau)$ with respect to $k$ to obtain the corresponding contribution to $\Expect{\phi_\ssC^2(\tau)}$. Keeping in mind that $G_\ssC^{(0)}(k,\tau) \propto (k \tau)^{2\epsilon_0}/k^3$, the required integral is
\be \label{FTint0}
 \int \frac{k^2 \exd k}{k^3} (k \tau)^{2\epsilon_0}
 \Bigl[ \ln( -k \tau) \Bigr]^L
 \propto \left( \frac{1}{\epsilon_0} \right)^{L+1} \Bigl[1 +
 \cO(\epsilon_0) \Bigr] \,,
\ee
and so the $m_0^2$-dependence of the contribution of the $L$-loop chain graph is\footnote{We thank Don Marolf and Ian Morrison for helpful correspondence on this point.}
\be \label{posm0}
 {\Expect{ \phi^2_\ssC(\tau)}}_\ssL \propto {\Expect{
 \phi^2_\ssC(\tau)}}_0   \left( \frac{\lambda H^4}{4 \pi^2 m_0^4}
 \right)^{L} \,.
\ee
At face value this indicates the edge of the perturbative domain lies at $m_0^2 \simeq \sqrt\lambda \; H^2/2\pi$, where the contribution of fluctuations to the mass begin to compete with $m_0^2$.

Notice, however, that because the dominant contribution to the Fourier transform comes from the $k \simeq 0$ limit of integration, the mass dependence of eq.~\pref{posm0} is exquisitely sensitive to the small-$k$ form of $G_\ssC(k,\tau)$, which the above graphs shows has an expansion in powers of $(\lambda/\epsilon_0) \ln(-k \tau)$. Yet we know that for sufficiently small $k$ this expansion breaks down because the logarithm systematically competes with the additional power $\lambda/\epsilon_0$. A better approximation for $G_\ssC(k,\tau)$ at small $k$ can be found by resumming the leading logarithms to obtain a result that is a series in $\lambda/\epsilon_0$, without accompanying factors of $\ln(-k \tau)$.

This resummation can be done \cite{us} using the dynamical renormalization group (DRG) \cite{drg}, by recognizing that because the integrand is a function of $k \tau$, small $k$ is related to large $\tau$. The breakdown of the perturbative expansion at small $k$ can therefore alternatively be regarded as being a breakdown due to the presence of the secular growth in $\tau$; a breakdown that the DRG is designed to resum. A better estimate of the small-$m$ behaviour of $\langle \phi^2_\ssC(\tau) \rangle$ can therefore be obtained by using the DRG-improved late-time approximation to $G_\ssC(k,\tau)$ that also improves its small-$k$ asymptotics.

The result of this resummation \cite{us} is to replace the expansion
\be
 G_\ssC(k,\tau) = G_\ssC^{(0)}(k,\tau) \left\{ 1 +
 \frac{\lambda}{6(2\pi)^2\epsilon_0}
 \left( \frac{\mu}{H} \right)^{2\epsilon_0} \ln(-k\tau)
 + \cO \left[ \left( \frac{\lambda}{\epsilon_0} \ln(-k \tau) \right)^2
 \right] \right\} \,,
\ee
with the DRG-improved result
\be
 G_\ssC(k,\tau) = G_\ssC^{(0)}(k,\tau) (-k\tau)^{ 2\delta}
 \left[ 1 +\cO \left( \frac{\lambda}{\epsilon_0} \right) \right]\,,
\ee
with
\be
 \delta = \frac{\lambda}{12(2\pi)^2 2 \epsilon_0} \,.
\ee
The key observation is that because $G_\ssC^{(0)}(k,\tau) \propto (H^2/k^3) (-k\tau)^{2\epsilon_0}$, the small-$k$ behavior of the DRG-resummed contribution to $G_\ssC(k,\tau)$ is equivalent to what would arise from a small shift $\epsilon_0 \to \epsilon = \epsilon_0 + \delta$. Equivalently, this corresponds to a shift of the scalar mass $m_0^2 \to m^2 = m_0^2 + \delta m^2$ with \cite{us}
\be \label{massshift}
 m^2 = m_0^2 + \frac{3 \lambda H^4}{16\pi^2 m_0^2} \,,
\ee
where $\epsilon = m^2/3H^2$.

Fourier transforming this more accurate depiction of the small-$k$ limit in $G_\ssC(k,\tau)$ to obtain $\Expect{\phi^2_\ssC(\tau)}$ then gives the following integral,
\be
 \langle \phi^2_\ssC(\tau) \propto H^2
 \int \frac{k^2 \exd k}{k^3} (k \tau)^{2\epsilon}
 \propto \frac{H^2}{\epsilon}  \,,
\ee
instead of eq.~\pref{FTint0}. This reproduces the above series in powers of $\lambda H^4/m_0^4$ once expanded using $1/\epsilon = 1/\epsilon_0 - \delta/\epsilon_0^2 + \cdots$. This shows that it is the scale $m^2$ that cuts off the IR divergences in $\langle \phi^2_\ssC(\tau) \rangle$, suggesting the utility of reorganizing the perturbative expansion so that it is the mass $m^2$ rather than $m_0^2$ that appears in the unperturbed lagrangian.

The corrections to the DRG-resummed form are of order
\be
 \frac{\lambda}{\epsilon_0} \propto \frac{\lambda H^2}{ m_0^2} \,,
\ee
suggesting that the boundary of the reorganized perturbation theory lies at $m_0^2 \simeq \lambda H^2/4\pi^2$, rather than when $m^2 \simeq \sqrt\lambda \; H^2/4\pi^2$, as was found above. We regard it to be an open question whether the perturbative problems that arise here can themselves  be resummed in a controllable fashion. The burden on any proponents of resummation is to show that the resummed graphs capture {\em all} of the leading $1/m$ behaviour, in the regime of interest. Experience with finite-temperature systems argues that although resummation may be possible for some regimes (like $m_0^2 \simeq 0$) this need not imply it can always be done (such as when $m^2 \simeq 0$). What is not yet clear is whether it is possible in de Sitter space to reach the regime $m^2 \simeq 0$, since $m^2$ does not pass through zero for any value of $m_0^2$, at least within the domain of validity for which eq.~\pref{massshift} holds.

\section{Discussion and Conclusions}

The body of this paper argues that an $L$-loop contribution to a
correlation function for a scalar field in de Sitter spacetime
with $\lambda \, \phi^4$ self-interactions carries a systematic
factor of $(\lambda H^2/4\pi^2 m_0^2)^{L}$, indicating a fundamental
breakdown of semiclassical methods once $m_0 \lsim \sqrt\lambda \;
H/2\pi$. The origin of the breakdown of perturbative methods is the infrared-singular behavior of these graphs which arises due to the large extra-Hubble fluctuations experienced by very light scalars in de
Sitter space. These fluctuations dominate the contributions to
correlation functions, invalidating the semiclassical
approximation which is at its heart a mean-field description.

This story is qualitatively similar to what happens in finite temperature field theory, although the power of coupling constant that defines the boundary of the semiclassical region differs. In the finite temperature case, the variance of the field, $\Expect{\phi^2}$, goes like $T^ 2$ while in the de Sitter case we find $\Expect{\phi^2}\sim H^4/m^2$. This difference means that the mass, $m_{\rm dyn}$, which is comparable to the corrections to $m$ scales differently with $\lambda$ in these two cases: for the thermal case $m_{\rm dyn}^2 \propto \lambda T^2$, while for a de Sitter background $m_{\rm dyn}^2 \propto \lambda^{1/2} H^2$.
It also may mean that the structure of de Sitter space precludes access to the regime $m^2 \simeq 0$ for any choice of $m_0^2$, unlike for finite-temperature systems, see Figure (\ref{fig:range}).
However, as is shown above, the two theories have a similar perturbative structure, whose relation is sketched in Figure (\ref{fig:chart}).

\FIGURE[ht]{
\epsfig{file=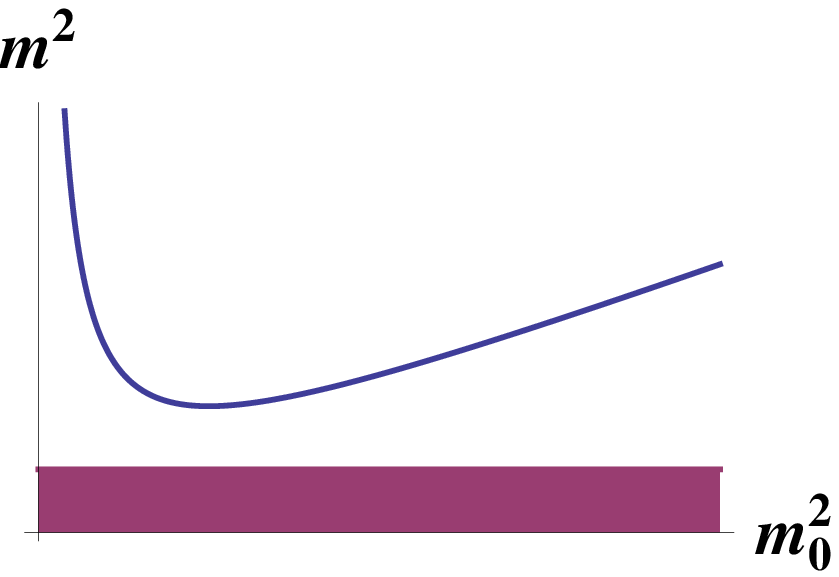,angle=0,width=0.5\hsize}
\caption{A plot of $m^2$ vs $m_0^2$. The horizontal
band represents the regime $m^2 < \cO(\lambda H^2/4\pi^2)$ while the curve is given by Eq. (\ref{massshift}). It would be interesting to investigate the case of negative bare mass, which we do not display here, further} \label{fig:range}}

A natural question to ask about this perturbative breakdown at small $m^2$ is whether it can be resummed (like for $m_0^2 \ge 0$ at finite temperature), or whether it reveals a complete breakdown of expansions based on powers of $\lambda$ (like when $m_0^2 < 0$ is chosen so that $m^2 = 0$ at finite temperature).

\FIGURE[ht]{
\epsfig{file=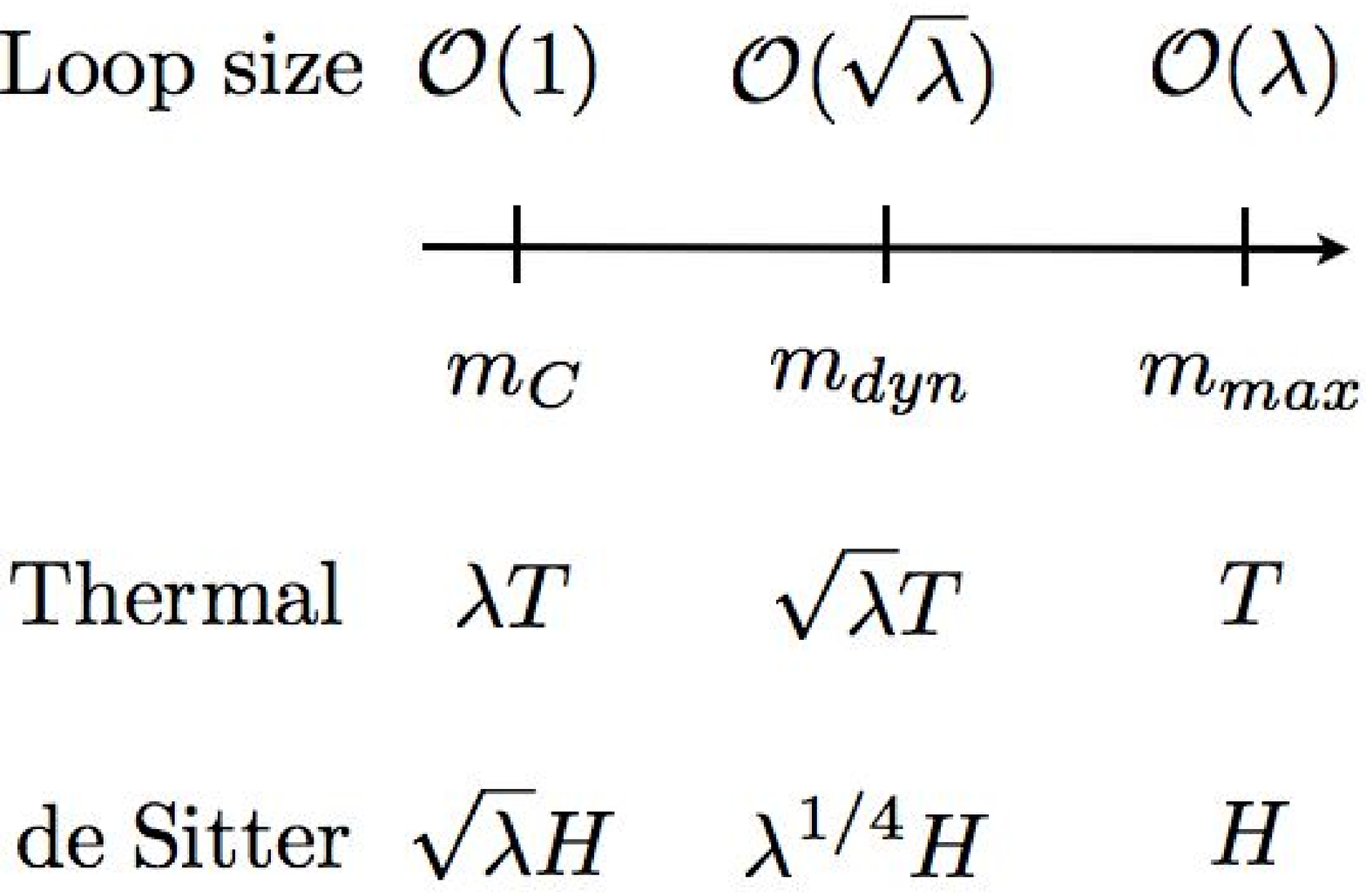,angle=0,width=0.6\hsize}
\caption{A comparison of the behavior of loop corrections for a
scalar field with a quartic interaction $\lambda\, \phi^4$ in a
thermal background or de Sitter space as a function of mass. The
masses labelled on the chart are, from left to right, the mass
below which perturbation theory breaks down, the dynamical mass
generated in the event that the zero temperature or flat space
mass was zero, and the maximum mass the field can have and still
receive thermal/de Sitter fluctuations.} \label{fig:chart} }

Others have argued that there exist semiclassical methods that capture the leading infrared logs. For secular growing logs, one proposal solves the late-time physics using the classical equation of motion \cite{MS}, in a similar way to the $\delta N$ formalism. Another point of view uses a stochastic approach to inflation \cite{superhubble}, which is argued to capture (and resum) the leading logs, by generating a dynamical mass \cite{SY}. Although such a stochastic approach goes beyond mean field, it is not yet clear what combination of small parameters control the approximations made in its derivation.

{}From the point of view of the arguments made here, derivative
couplings are not as dangerous as are those of the scalar
potential. This is because the momentum dependence of these
couplings tends to ameliorate any IR divergence the graph would
have otherwise had. This means that massless Goldstone bosons on
de Sitter space would {\em not} suffer from the same breakdown of
perturbation theory as does the $\phi^4$ model considered here.

We would expect the self-interactions of massless gravitons to be
similarly benign so long as these are derivative couplings,
leading us to expect no perturbative breakdown for pure gravity on
de Sitter space. This expectation seems to be borne out by ref.~\cite{DB}
and ref.~\cite{GS} (the latter appeared just as our paper neared
completion). These authors study one-loop infrared divergences for
pure gravity and gravity coupled to scalars in de Sitter and
slow-roll spacetimes, but find that IR divergences cancel in the
absence of the self-interactions of the scalar potential (which
appear in their calculations as slow-roll parameters). Based on
the power-counting arguments presented here we expect this to
continue to be true at higher loops, with the scalar
self-interactions being the most dangerous in the infrared.

It is interesting to consider in this light what implications our result might have for inflationary cosmologies. A complication in directly extracting these for simple single-field models is our neglect of classical evolution of the background scalar field and metric, due to our use of a simple de Sitter background. Because a homogeneous evolving scalar field can be used to define a notion of cosmic time, many of its effects can be gauged away. We therefore expect a naive application of the above arguments to simple models are likely to cancel from gauge invariant quantities like the curvature fluctuations of physical interest for cosmology. Nevertheless, it may be possible to have important infrared effects appear in curvature correlations in multi-field models, particularly those involving nontrivial post-inflationary dynamics (such as curvaton models). We leave for future work the detailed question of whether and how the infrared effects we find above `propagate' into late-time curvature perturbations.

In the remainder of this section we put these issues aside, and ask what the condition $m^2 > \lambda H^2/4\pi^2$ implies for the parameters of a single scalar field described by a quartic potential
\be
 V = V_0 + \frac12 \, m^2
 \phi^2 + \frac{1}{4!} \, \lambda\, \phi^4\,.
\ee
For this model we ask how the condition $M^2 \gg \lambda (H/2\pi)^2$, where $M^2 = V'' = m^2 + \frac12 \lambda \, \phi^2$, compares to other conditions to which inflationary models must be subject. To see what this implies for
potential inflationary applications, consider two extreme cases: $(i)$
$\phi$ so large that $V \simeq \frac{1}{24} \, \lambda \, \phi^4$;
and $(ii)$ $\phi$ small enough that $V \simeq V_0$. (This last
example can be regarded as part of a model of hybrid inflation
\cite{hybrid}, with inflation ending as another field starts to
roll as its $\phi$-dependent squared-mass goes negative.)

\subsubsection*{Large-field inflation}

For the large-$\phi$ regime we have $V \simeq \frac{1}{24} \,
\lambda \, \phi^4$, and so $H^2 = V/3M_p^2 \simeq \lambda \,
\phi^4/72 M_p^2$. The slow-roll parameters are
\be
 \varepsilon := \frac12 \left( \frac{M_p V'}{V} \right)^2
 \simeq 4 \left( \frac{M_p}{\phi} \right)^2
 \quad \hbox{and} \quad
 \eta := \frac{M_p^2 V''}{V}
 \simeq 12 \left( \frac{M_p}{\phi} \right)^2  \,,
\ee
so the edge of the inflationary regime is $\phi_\SR /M_p \simeq
\O(1)$. For $\phi$ larger than this classical evolution satisfies
\be
 \dot\phi \simeq \frac{V'}{3H} \simeq \frac{\frac16 \, \lambda \,
 \phi^3}{\sqrt\lambda \; \phi^2/2\sqrt2 \; M_p} \simeq
 \frac{\sqrt{2}\,}{3} \, \sqrt\lambda \; M_p \phi \,.
\ee
Once $\dot\phi$ becomes smaller than $H^2$ fluctuations dominate
classical evolution and inflation becomes eternal, which in this
case occurs when $\frac{\sqrt{2}\,}{3} \sqrt\lambda \; M_p \phi <
\lambda \, \phi^4/72 M_p^2$, or $\phi^3 > \phi_\EI^3 \simeq 24
\sqrt2 \; M_p^3/\sqrt\lambda$.

How do the boundaries of the semiclassical approximation compare
to these? There are two criteria to be satisfied. First, control
over the low-energy approximation that underlies the gravitational
loop expansion requires $V \ll M_p^4$, or $\phi^4 \ll \phi_\HE^4
\simeq 24 M_p^4/\lambda$. We have seen in previous sections that
the $\lambda$ loop expansion fails unless $M^2 \gg \lambda \,
H^2/4 \pi^2$ or $\frac12 \, \lambda \, \phi^2 \gg \lambda^2
\phi^4/288 \pi^2 M_p^2$, and so $\phi^2 \ll \phi^2_\BD \simeq (12
\pi M_p)^2/\lambda$. Since $\phi_\SR/M_p \simeq \O(1)$,
$\phi_\EI/M_p \simeq \O(\lambda^{-1/6})$, $\phi_\HE /M_p \simeq
\O( \lambda^{-1/4})$ and $\phi_\BD /M_p \simeq \O(
\lambda^{-1/2})$ we have
\be
 \phi_\SR \ll \phi_\EI \ll \phi_\HE \ll \phi_\BD \,,
\ee
and so the condition $M^2 > \lambda (H/2\pi)^2$ is parametrically
weaker than the condition $M_p^4 > V$. The fact that eternal
inflation can occur before a total breakdown of perturbation
theory is as expected.

\subsubsection*{Hybrid inflation}

Consider next a hybrid model \cite{hybrid}, involving two scalar
fields, $\phi$ and $\chi$, interacting through the potential
\be
 U(\phi,\chi) = \frac{1}{4} \,\zeta \,( \chi^2 - v^2 )^2 +
 \frac{g^2}{2} \, \chi^2 \phi^2 + \frac12 \, m^2
 \phi^2 + \frac{1}{4!} \, \lambda\, \phi^4\;.
\ee
In this model the fields start in the trough defined by $\chi =
0$, with $\phi$ large and rolling towards smaller values subject
to the effective potential
\be
 V = V_0 + \frac12 \, m^2
 \phi^2 + \frac{1}{4!} \, \lambda\, \phi^4\,,
\ee
with $V_0 = \frac{1}{4!} \, \zeta \, v^4$. This roll continues
until
\be
 \phi = \phi_\SR = \frac{\sqrt\zeta \, v}{g} \,,
\ee
at which point the $\chi$ mass, $\mu^2 = - \zeta \, v^2 + g^2
\phi^2$, becomes negative, allowing $\chi$ to evolve quickly
towards the absolute minimum at $\chi = v$ and $\phi = 0$.

The dynamics of the inflaton in this model is governed by the same
potential considered earlier, but our interest now is in the
small-field regime, for which $m^2 \gg \frac12 \, \lambda \,
\phi^2$ and $V \simeq V_0$ (and so $H^2 \simeq V_0/3M_p^2$). We
assume parameters are chosen to keep $\phi_\SR$ small enough to
ensure $\chi$ remains zero well into this regime. In this case the
slow-roll parameters are
\be
 \varepsilon \simeq \frac12 \left( \frac{m^2 M_p \phi}{V_0}
 \right)^2  \quad \hbox{and} \quad
 \eta \simeq \frac{m^2 M_p^2}{V_0}  \,,
\ee
so $2 \varepsilon \simeq \eta^2 (\phi/M_p)^2$. Clearly inflation
only requires $m^2 M_p^2 /V_0 \ll 1$, since the condition $\frac12
\, \lambda \phi^2 \ll m^2$ automatically ensures $\phi \ll
M_p/\eta$. Inflation ends once $\phi$ reaches $\phi_\SR =
\sqrt\zeta \; v/g$.

In this case the validity of the $\lambda$ loop expansion requires
$m^2 \gg \lambda \, H^2/4 \pi^2 \simeq \lambda V_0/12 \pi^2
M_p^2$, or
\be
 \eta \simeq \frac{m^2 M_p^2}{V_0} \gg \frac{\lambda}{12 \pi^2} \,.
\ee
Notice that this lower bound to the slow-roll parameters is not
simply a naturality condition, corresponding to a regime for which
small loop corrections dominate the smaller classical potential.
Rather, in this regime the problem is not fixed simply by
including one- or two-loop corrections to $V$. Instead it is the
entire loop expansion itself that fails. It would be of great
interest to see whether the semiclassical criterion plays any role
in more general inflationary contexts.

\section*{Acknowledgements}

We would like to thank Don Marolf, Ian Morrison, Arvind Rajaraman, Tony Riotto, David Seery, Leonardo Senatore and Andrew Tolley for useful conversations.
This research has been supported in part by funds from the Natural
Sciences and Engineering Research Council (NSERC) of Canada (CB),
and the US Department of Energy (RH), through Grant  No.
DE-FG03-91-ER40682. Research at the Perimeter Institute is
supported in part by the Government of Canada through Industry
Canada and by the Province of Ontario through the Ministry of
Research and Information (MRI).

\appendix

\section{Power counting in k-space}\label{example}

In this appendix we examine the leading infrared behaviour of several some Feynman graphs, to follow how the integrations over loop momenta, $p$, and vertex times, $\tau_i$, reproduce the logarithmic dependence on $\Lambda_\IR$. We illustrate many of the subtleties that arise when trying to do power counting in $k$ space in this context.

\begin{figure}[h]
\begin{center}
\includegraphics[width=0.5\textwidth,angle=0]{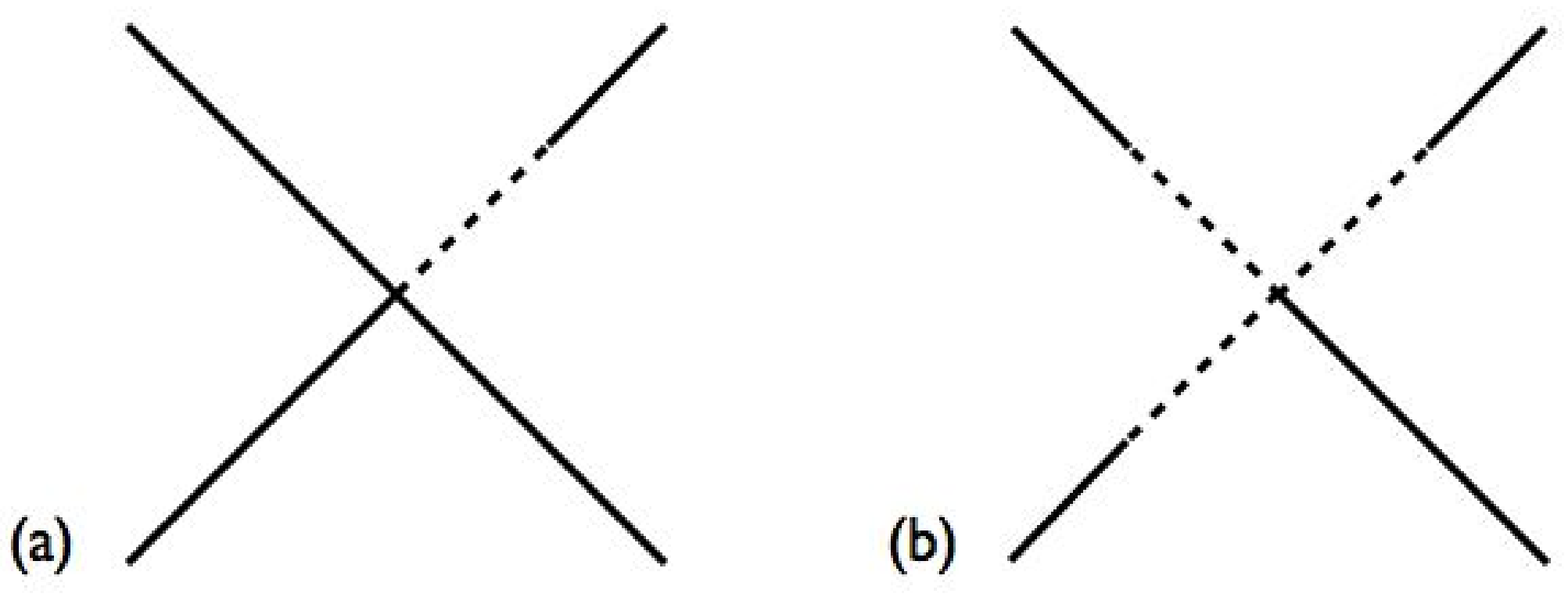}
\caption{The diagram on the left (a) is $E_C = 3$ $V_C =1$ and the diagram on the right (b) has $E_C =1$ and $V_\Delta = 1$. }
\label{fig:treelevel}
\end{center}
\end{figure}

For most diagrams the integral over time either converges well in the infrared ({\em i.e.} gives terms of order $ \mathcal{O}(1) + \mathcal{O}(-k\tau)$) or diverges logarithmically, like $\ln( -k\tau)$ \cite{us}. Weinberg has proven that the log divergence is the worse divergence you can get under a certain set of assumptions that are satisfied by our $\lambda\phi^4$ model \cite{Wbg}. Nevertheless, for some diagrams the time integral give important powers of external momentum which one must keep track of to get the power counting right.  To see this let us first look at the tree level calculation of $\Expect{\phi_C^4}$. There are two diagrams that contribute to this correlation function at tree level --- see Fig.~(\ref{fig:treelevel}).
For simplicity and illustrative purposes, we take all external momenta and time to be equal to $k$ and $\tau$. The first diagram (diagram (a) with $E_C = 3$) gives
\ba\label{diagrama}
 \Expect{\phi_C^4} &\supset& \frac{H^4}{k^9}\int_{-1/k}^{\tau} \frac{d\tau'}{\tau'^4}  (\tau^3 -\tau'^3)\nonumber\\
 &\supset & \frac{H^4}{k^9} \left(\ln(-k\tau) + \O((-k\tau)^n)\right)
\ea
with $n > 0$. The integration limits need to be explained; the upper limit arises from the $\theta(\tau -\tau')$ in the retarded propagator. By causality, the interaction occurred at any time in the past of $\tau$. But if the interaction occurs so far in the past that any of the external momenta come inside the horizon, $-k\tau' \sim 1$, then the corresponding propagator starts oscillating very quickly and the integral will in general vanish: $\int_{-\infty}^{-1/k} (\cdots) e^{-ik\tau'} \sim 0$.
The second diagram (diagram b) in Fig.~(\ref{fig:treelevel}) on the other hand goes like
\ba
 \Expect{\phi_C^4} &\supset & \frac{H^4}{k^3}\int_{-1/k}^{\tau} \frac{d\tau'}{\tau'^4}  (\tau^3 -\tau'^3)^3\nonumber\\
 &\supset & \frac{H^4}{k^3} \frac{1}{k^6}\left(\O(1) + \O((-k\tau)^n)\right)
\ea
We see that the time integral give the $1/k^6$ that was na\"ively missing from the second diagram.  Note that it could be that the $\O(1)$ factors cancel and the diagram would then be $\sim \frac{\tau^A}{k^B}$ such that $A + B = 9$. Since we are working with $-k\tau \ll1$, the diagram would be much smaller than expected. Our power counting is only good to keep track of the leading piece and it assumes no special cancellations. This assumption is completely justified given the absence of special symmetry.

In loop diagrams, we have identified at least three additional complications that arise from the integral over internal momentum $p$ and from the entangled time integrals. The first subtlety occurs because the momentum of the virtual particle is integrated over all values and when $p> k$, the lower cutoff of the time integral becomes $-1/p$ instead of $-1/k$. This is because the virtual particle stops oscillating at a later time, or leaves the horizon at a later time. Coming in from $\tau' \sim -\infty$, we find that once we reach the time $\tau' \sim -1/k$, the virtual propagator is still oscillating and the time integral keeps integrating to 0 until we reach the time $\tau'\sim -1/p$.
The second subtlety is that the retarded propagator imposes some kind of ordering on the internal vertex times $\tau_i > \tau_k > \cdots$ such that the upper bound of some integrals are variable of other integrals. Finally the third subtlety arise from the fact that the momentum integral may depend on time as well. This could occur for example if one deals with infrared effects with a beginning to inflation or if one uses an explicit IR cutoff.

All these subtleties make the power counting very complicated in $k$ space and this is why we limited ourselves to the real space expression in the main text Eq.~(\ref{powergeneralposn}).
Still, progress can be made to keep track of the leading infrared piece if we make the following set of assumptions.
We will take all time integral to stop at  some external time $\tau$ and take all internal momenta to be smaller than external momenta $p <k $. We can integrate all internal time $\tau_i$ to external times $\tau_a$.  Finally, the IR physics is a small mass and no time dependence is introduced in the integral over momenta from infrared effects.
Applying these simple rules to do the power counting of the time integrals in Eq.~(\ref{powergeneral}), we get
\ba
 {\rm time} & \sim & \left[\int\frac{d\tau_i}{\tau_i^4}\right]^{V_C+V_{\Delta}}  \left[\theta(\tau-\tau^{\prime})(\tau^3-\tau^{\prime3})
 \right]^{E_R^{(1)}+E_R^{(2)}}\left[\frac{\theta(\tau_i-
 \tau^{\prime})}{3}(\tau_i^3-\tau^{\prime3})\right]^{I_R}\nonumber\\
 &\sim& \left[\theta(\tau_a -\tau_b) (\tau_a^3)\right]^{E_R^{(2)}} \left[\int_{-1/k}^{\tau_c} d\tau_i\right]^{V_C+V_{\Delta}} (\tau_i)^{3E_R^{(1)} + 3I_R - 4V_c -4V_\Delta} + \cdots
\ea
where we have factorized the $\ERo$ part and we have kept only one term (where all $\tau'$ are $\tau_i$) typical among many. This gives
\be
 {\rm time} \sim \left[\theta(\tau_a -\tau_b) (\tau_a^3-\tau_b^3)\right]^{E_R^{(2)}} \frac{1}{k^{6V_\Delta} } \left( \O(1) + \O((-k\tau)^n\right)
\ee
where we combined terms to generate the full correct $\ERo$ propagator (it has to), $n>0$ and $\O(1)$ could be a $\ln(-k\tau)$ as well. In summary, the main effect of the time integrals is to give rise to $1/k^{6V_\Delta}$ and note that the power counting does not keep track of the $\ln(-k\tau)$.

\begin{figure} [ht]
\begin{center}
\includegraphics[width=0.5\textwidth,angle=0]{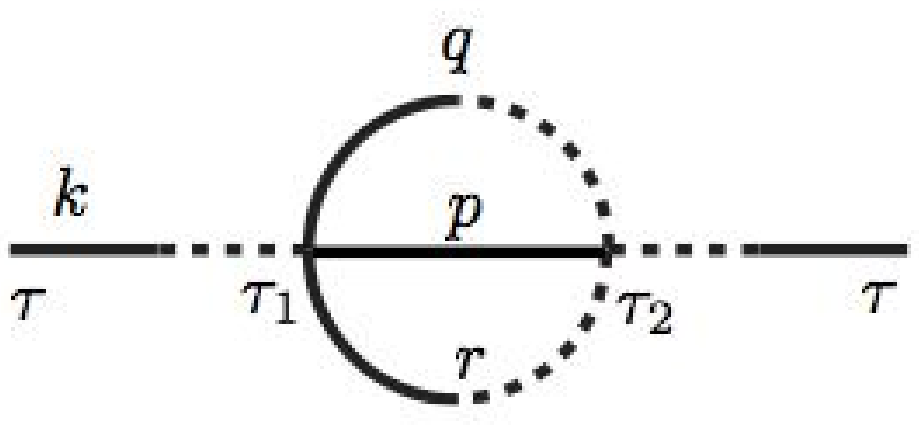}
\caption{The time are from left to right $\tau$, $\tau_1$ and $\tau_2$. The incoming momentum is $k$. The internal momentum from top to bottom are $q$, $p$ and $r = |\vec{k} - \vec{q} - \vec{p}|$.} \label{fig:sunset}
\end{center}
\end{figure}

To illustrate further the complication of doing these loop diagrams, lets look at the following sunset diagram (see Fig \ref{fig:sunset}).
This is a 2-loop correction to $\Expect{\phi_C\phi_C}$ which goes like
\be\label{sunset}
\int \frac{d^3p}{p^3} \int d^3 q \int \frac{d\tau_1}{\tau_1^4} \int \frac{d\tau_2}{\tau_2^4} \theta_{\tau,1} \theta_{1,2} \theta_{2,\tau} (\tau^3-\tau_1^3) (\tau_1^3-\tau_2^3)^2 (\tau_2^3-\tau^3)
\ee
where we use the notation that $\theta_{\tau, 1} = \theta(\tau - \tau_1)$.  The theta functions give the following hierarchy $\tau > \tau_1 >\tau_2$ and all the propagators are only valid given a series of condition
\be
-\{p,q,k\} \{\tau,\tau_1,\tau_2\} \ll 1
\ee
There are multiple terms (or regime) to consider but the most important ones are when $p$ is in the IR. For example, in the regime where $p, q \ll k$,
the external momentum provides the most stringent constraints and so we have that $\{\tau,\tau_1,\tau_2\} \gg -1/k$. So a typical term in Eq.~(\ref{sunset}) goes like
\ba
 &\sim&\int_{0}^k \frac{d^3p}{p^3} \int_0^k d^3 q   \int_{-1/k}^{\tau} \frac{d\tau_1}{\tau_1^4} \int_{-1/k}^{\tau_1} \frac{d\tau_2}{\tau_2^4} \tau_2^9\tau_1^3\\
&\sim& \int_{0}^k \frac{d^3p}{p^3} \int_0^k d^3 q \left(\frac{1}{k^6} + \mathcal{O}((-k\tau)^n)\right)
\ea
where as promised the result of the time integral is to give $1/k^{6V_\Delta}$ with $V_\Delta =1$ above. There are other terms of the same order as this one and others which are smaller (usually by powers of $-k\tau$).
The integral over internal $q$ will give $k^3$ such that the overall correct scaling of $1/k^3$ for $\Expect{\phi_C^2}$ is recovered while the $p$ integral is infrared divergent. If we use a massive propagator instead of the massless one, we would get a $\frac{1}{\epsilon} \sim \frac{H^2}{m^2}$ or in terms of $\Lambda_{IR}$ a log divergence.
So, barring cancelations the sunset diagram (Fig. \ref{fig:sunset}) scales like
\be
 \cA \sim \delta^3(\sum_i \vec{k}_i )\frac{H^2}{k^3} \frac{\lambda^2}{\epsilon}
\ee
where we include the external legs and the overall $\delta^3$ function. This two-loop diagram is therefore subdominant to the equivalent two-loop chain diagram by one power of $1/\epsilon$.

\end{document}